\documentclass{iopconfser}
\usepackage{orcidlink}
\usepackage{hyperref}

\usepackage{graphicx}
\usepackage{dcolumn}
\usepackage{bm}
\usepackage{wasysym}
\usepackage{hyperref}
\usepackage[mathlines]{lineno}
\usepackage{xcolor}
\usepackage{dsfont}

\usepackage{booktabs}

\usepackage{gensymb}
\usepackage{enumitem}
\usepackage{multirow}
\usepackage{booktabs}
\usepackage{array}
\newcolumntype{P}[1]{>{\centering\arraybackslash}p{#1}}
\usepackage{amssymb}
\usepackage{adjustbox}
\usepackage{orcidlink}
\usepackage{float}

\usepackage[normalem]{ulem}

\newcommand{\iu}{\mathrm{i}\mkern1mu}
\newcommand{\du}{\mathrm{d}}

\begin{document}

\title{Perturbations of Solitonic Boson Stars: Nonlinear Radial Stability and Binding Energy}

\author{Gareth Arturo Marks$^{1}$ \orcidlink{0009-0003-3160-9337}} 

\affil{DAMTP, Centre for Mathematical Sciences,
University of Cambridge, Wilberforce Road, Cambridge CB3 0WA, UK}

\email{gam54@cam.ac.uk}

\begin{abstract}
We study the nonlinear radial stability of boson stars with a solitonic potential across the entire parameter space, focusing especially on families of solutions that support ultracompact models on the perturbatively stable branch. Using a dimensional reduction of the CCZ4 formulation of numerical relativity, we dynamically evolve these models with both internal and external perturbations. We find in particular that there are perturbatively stable models with positive binding energy that do not effectively disperse even under explicit perturbations, challenging the conventional wisdom that negative binding energy is a necessary condition for the dynamical stability of boson stars and other compact objects.
\end{abstract}

\section{Introduction}
Since the introduction of Einstein's general relativity, a large number of exotic compact object (ECO) models have been proposed, responding to issues ranging from the unknown nature of dark matter to theoretical problems with classical black holes such as the information paradox \cite{Mathur:2009ip}. Boson stars (BSs) are a simple and numerically tractable example of an ECO, arising from the coupling of general relativity to a massive complex scalar field. After their original proposal by Kaup \cite{Kaup:1968} as well as Ruffini and Bonazzola \cite{Ruffini:1969qy}, BSs received attention in a variety of roles, including as candidates for dark matter haloes \cite{Sin_1994, Schive_2014} and black hole mimickers \cite{Torres:2000dw, Guzman:2005bs, Amaro_Seoane_2010}. Binary mergers of BSs have also been studied as a potential source of gravitational waves \cite{Bezares_2022, Helfer_2022, Evstafyeva:2023kfg}, some of which may be degenerate with binary black hole signals \cite{Evstafyeva:2024qvp}.

An interesting class of BS solutions can be obtained by using the so-called \textit{solitonic} potential, which significantly raises the compactnesses achievable in the perturbatively stable regime \cite{Lee:1987, Collodel:2022jly}. Indeed, these families contain \textit{ultracompact objects} (UCOs), solutions so compact that they support a pair of light rings, one stable and one unstable \cite{Cunha:2017qtt,Cunha:2020azh}. Solitonic BSs have therefore been used in previous numerical studies aimed at determining the efficiency of a conjectured instability related to the presence of a stable light ring \cite{Keir:2014oka, Cunha:2022gde, Marks:2025,Marks_2025_B, Evstafyeva:2025mvx}. 

If BSs are to be considered as viable astrophysical objects, their long-term dynamical stability is a key question. Previous numerical studies of BS evolutions have been made, beginning with the pioneering work of Seidel and Suen \cite{Seidel_Suen_1990,Balakrishna_1998} in spherical symmetry and continuing with full 3+1 evolutions \cite{Guzman_2004} and a careful study of the dynamical fate of perturbatively unstable solutions \cite{Guzman_2009}. This has also been compared to the results of radial perturbation theory, which in the case of \textit{mini} BSs corresponding to a simple massive Klein-Gordon potential, predicts a single radially stable branch \cite{Gleiser:1988ih, Kain:2021rmk}. The conclusion drawn by this work is that linear radial perturbation theory is sufficient to determine the nonlinear stability of mini BSs both in and out of spherical symmetry, with models on the perturbatively stable branch always exhibiting long-term dynamical stability and models on the unstable branch always failing to do so. Furthermore, for unstable models, the character of the instability depends on the sign of the \textit{binding energy}-- a quantity measuring the work required to assemble the BS configuration. Unstable models with negative binding energy migrate back to the stable branch or collapse to BHs depending on the type of perturbation added, while those with positive binding energy disperse, the scalar matter escaping to spatial infinity. Similar results have been seen in the case of the recently-proposed $\ell$-boson stars, spherically symmetric solutions characterized by a superposition of scalar fields which individually carry nonzero angular momentum \cite{Alcubierre_2018, Alcubierre_2019}.

In this work, we study the dynamical evolutions of solitonic BSs in spherical symmetry under explicit perturbations, including the black hole mimicker candidates previously evolved in Ref.~\cite{Marks:2025} with no evidence of a light-ring instability. We draw particular attention to the presence of perturbatively stable models with positive binding energy in these families. The reasonable heuristic that negative binding energy is necessary for stability appears frequently in the literature \cite{Guzman_2009, Hartmann_2010, Franzin_2024}; we therefore aim to test the robustness of this heuristic by determining whether radial perturbations can trigger an instability to fission. We employ Planck units $c=G=\hbar=1$, but keep the
scalar-field mass parameter $\mu$ which controls the scale invariance
of our BS spacetimes.

\section{Theory and Numerical Setup}
\begin{figure}[h]
    \vspace{-15pt}
    \includegraphics[width=\linewidth]{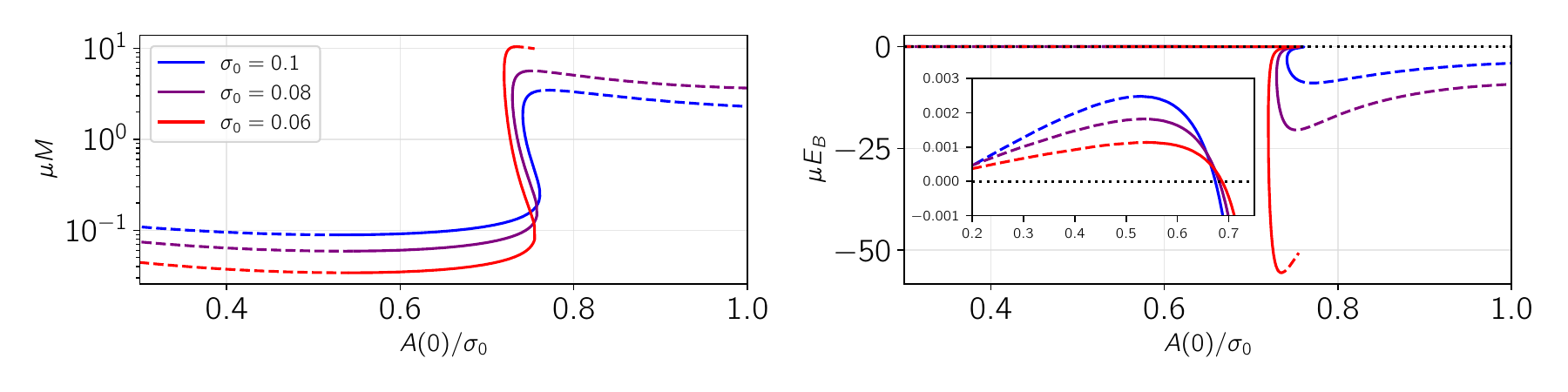}
    \vspace{-30pt}
    \caption{
    The ADM mass $M$ (left) and binding energy $E_B$ (right) against the central scalar-field amplitude
    $A_0$ for families of BS solutions with
    varying parameter $\sigma_0$. Solid lines indicate perturbative stability, while dashed indicate instability; see e.g. Ref.~\cite{Marks:2025} for details. The inset draws attention to the region in which linearly stable models with positive binding energy exist.
    \label{fig:M_and_E}}
\end{figure}

We take a minimally coupled, complex scalar field $\varphi$
with solitonic potential, described by the action
\begin{eqnarray}
  S=\int  \sqrt{-g} \left\{
  \frac{R}{16\pi}-\frac{1}{2}\left[
  g^{\mu\nu}\nabla_{\mu}\bar{\varphi}\,\nabla_{\nu}\varphi
  +V(\varphi)
  \right]
  \right\}\du^4 x,\quad
  V(\varphi) = \mu^2 |\varphi|^2 \left(
  1-2\frac{|\varphi|^2}{\sigma_0^2}
  \right)^2
  \label{eq:action}
\end{eqnarray}
with $\sigma_0$ a constant. Varying $S$, we obtain the Einstein-Klein-Gordon equations
\begin{eqnarray}
  && G_{\alpha\beta}=8\pi \,T_{\alpha\beta}\,,
  ~~~~~
  \nabla^{\mu}\nabla_{\mu}\varphi = \frac{\du V}{\du \bar{\varphi}}, \quad
   T_{\mu\nu}
  =
  \frac{1}{2} \nabla_{(\mu} \bar{\varphi}\,\nabla_{\nu)}\varphi
  -\frac{g_{\mu\nu}}{2}
  \left[
  g^{\alpha\beta}\nabla_{\alpha}\bar{\varphi}\,\nabla_{\beta}\varphi
  + V(\varphi)
  \right],
\end{eqnarray}
which we solve numerically with an ansatz of the form $\varphi(t,r)=A(r)e^{\iu \omega t},$ fixing the central amplitude $A_0$ and determining $\omega$ via a two-way shooting algorithm. In addition to the ADM mass $M$, there is a Noether charge $N$ associated with the $U(1)$ symmetry of the complex scalar field; the binding energy is then given by $E_B = M - \mu N$. Solutions with (positive) negative binding energy are interpreted as gravitationally (un)bound. In Fig.~\ref{fig:M_and_E} we plot the mass and binding energy against the central field amplitude for some solitonic families; for $\sigma_0 \lesssim 0.085$ these contain UCOs on the perturbatively stable branch.
We perform evolutions using {\sc
sphericalbsevolver}, a code introduced in Ref.~\cite{Marks:2025} that evolves dimensional reductions of the BSSN \cite{Baumgarte:1998te,Shibata:1995we} and CCZ4 \cite{Alic:2011gg} formulations of numerical relativity. Following Ref.~$\cite{Alcubierre_2019}$, we introduce perturbations to the real part of $\varphi$ and the imaginary part of the field momentum $\Pi_\varphi$ in such a way that the Noether charge is held constant to first order. We prescribe perturbations using a Gaussian profile of the form $\delta \varphi = a \exp\left((r - r_0)^2 / k^2\right), \; \delta \Pi_\varphi = -i\omega \delta\varphi / \alpha$ for real constants $a, r_0, k$ where $\alpha$ is the lapse.
\section{Results and Discussion}
We present a selection of characteristic results in Table \ref{tab:tab1}, divided into three categories: first, models on the stable and unstable (with both signs of $E_B$) branches (A); second, linearly stable models with $E_B > 0$ (B); third, ultracompact models (C). All stable evolutions are run for time at least $t = 10^5 M$ with no sign of instability. We see that our explicit perturbations introduce no departures from the predictions of the linear theory. In agreement with previous work, linearly unstable models collapse to BHs or migrate (in the case $E_B < 0$) or disperse (in the case $E_B > 0$), while linearly stable models remain stable in perturbed evolutions so long as the perturbation does not change their Noether charge enough to displace them to an unstable branch. This remains true for linearly stable models with $E_B > 0$: even with perturbations that remove scalar matter and hence further increase $E_B$ (cf. inset of Fig. \ref{fig:M_and_E}), we see no sign of an instability to fission. We conclude that, at least in spherical symmetry, the presence of positive binding energy does not necessarily introduce an effective instability. The claim that $E_B < 0$ is required for stability must therefore be regarded as a rule of thumb rather than a strict condition. Finally, we note that this picture remains true even for ultracompact configurations, which are located very close to the ends of the perturbatively stable branches, where one might expect nonlinear effects to become important. Nonetheless, we see no departures from the linear predictions. This is in contrast with the neutron star case, where nonlinearities in the radial oscillations have been found numerically to affect stability \cite{Sperhake_2001n}.
\begin{table}[h]
\centering
\footnotesize
\begin{tabular}{@{}|c|c|c|c|c|c|c|c|c|c|c|@{}}
\toprule
\textbf{Label} &$A_0$ &$\sigma_0$  &$\mu M$  & $\mu E_B$  &$C_{99}$   &$a$  &$k$  &$r_0$  & Lin. Stable?  & Outcome  \\ \midrule

 A1&0.09  &0.2   &0.271  &0.00236  &0.0403  &0.001  &0.1  &10.0  &No  &Unstable (disperse)  \\ 
 A2&0.17  &0.2  &0.713  &-0.334  &0.179  &0.005  &1.0  &3.0  &Yes  &Stable  \\ 
 A3&0.18  &0.2  &0.706  &-0.325  &0.190  &-0.001  &1.0 & 3.0  &No  &Unstable (collapse)  \\ 
 B1&0.035  &0.06  &0.0345  &0.00109  &0.00516  &-0.001  &1.0  &2.0  &Yes  &Stable  \\ 
 B2&0.045  &0.08  &0.0593  &0.00179  &0.00867  &0.001  &1.0  &3.0  &Yes  &Stable  \\ 
 B3&0.055  &0.1  &0.0893  &0.0245  &0.0131  &-0.001  &2.0  &0.0  &Yes  &Stable  \\ 
 C1&0.044  &0.06  &10.43  &-55.9  &0.328  &0.001  &0.5  &10.0  &Yes  &Stable  \\ 
 C2&0.045  &0.06  &0.06  &-52.2  &0.340  &-0.001  &1.0  &5.0  &No  &Unstable (collapse)  \\ 
 C3&0.06  &0.08  &5.65  &-20.5  &0.313  &-0.001  &0.5  &10.0  &Yes  &Stable  \\ 
\bottomrule
\end{tabular}
    \caption{
    Evolution results, showing  the central amplitude $A_0$, solitonic parameter $\sigma_0$, BS mass $M$, binding energy $E_B$, compactness $C_{99} = M / r_{99}$ with $r_{99}$ defined as the radius enclosing 99\% of the mass,  perturbation parameters $a, k, r_0$, whether the model is linearly stable, and the evolution outcome.
    }
\label{tab:tab1}
\end{table}

\begin{figure}[h]
    \includegraphics[width=\linewidth]{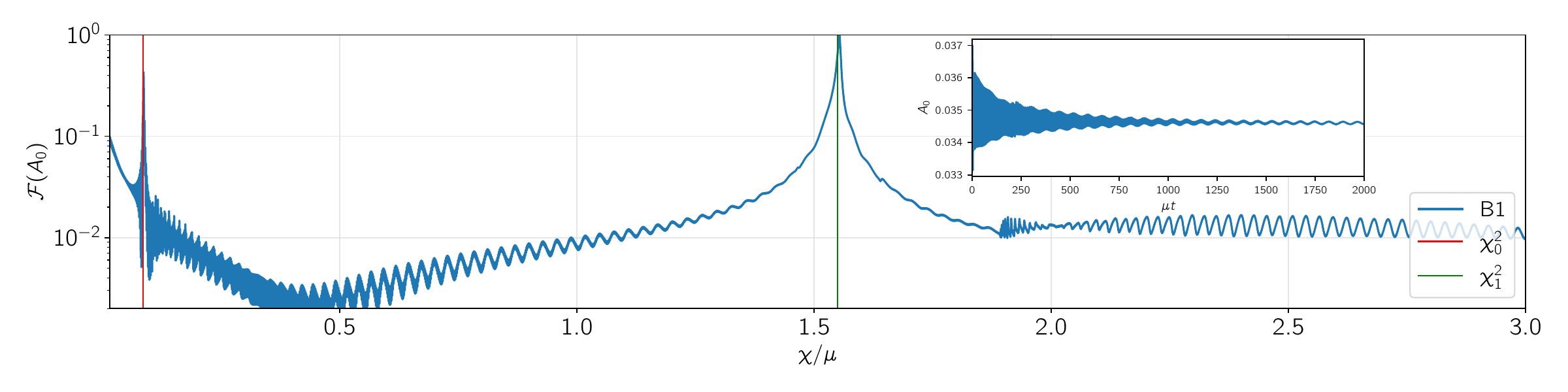}
    \vspace{-30pt}
    \caption{
    Power spectrum of the central amplitude oscillations for evolution B1, showing peaks at the fundamental frequency $\chi_0$ and first excited frequency $\chi_1$, represented by vertical lines. The inset displays $A_0$ in the time domain for reference.
    \label{fig:fluffy_frequencies}}
\end{figure}

In Fig. \ref{fig:fluffy_frequencies}, we show the power spectrum of the central amplitude for the B1 run. We see that the fundamental and first excited radial oscillation frequencies (see Ref.~\cite{Marks:2025} for the method to compute these), accounting for displacement due to the initial perturbation, agree with the two largest peaks. This result is generic, underscoring the point that linear theory predicts the dynamical evolution of these models very well. Furthermore, the time-domain evolution makes it clear that the perturbed BS settles down gradually into a stationary solution with positive binding energy.

\section{Conclusions}
Our results provide additional evidence that the radial stability of BSs does not depend significantly on nonlinear effects, including for solitonic models and in the ultracompact case. Furthermore, we have presented a concrete example of a model which shows robust nonlinear radial stability, despite having positive binding energy. We conjecture that such models will not generically form under gravitational collapse, but that the solitonic potential introduces an obstruction preventing transition to the energetically preferred state in which all scalar matter escapes to infinity. In future work, it would be useful to evolve these models out of spherical symmetry to determine whether non-radial modes permit additional dispersive instabilities. Such an instability might be surprising, given the lack of nonspherical instabilities present for ground-state non-rotating BSs in the literature.  It would also be interesting to repeat this analysis for a broader class of scalar potentials, to determine whether other models with degenerate vacuua, such as the axionic potential \cite{Siemonsen_2021}, display similar behaviour. Finally, the formation of solitonic BS models has received little attention, and we plan to investigate this in the future, especially in the case of UCOs.

\section*{Acknowledgments}
G.A.M. is supported by the Cambridge Trust at the University of Cambridge. I am grateful for useful discussions with Seppe Staelens, Tamara Evstafyeva, and Christopher Moore, and especially to my supervisor Ulrich Sperhake for his very helpful guidance in carrying out this research. We acknowledge support by the NSF Grant Nos.~PHY-090003,~PHY-1626190
and PHY-2110594, DiRAC projects ACTP284 and ACTP238, STFC capital
Grants Nos.~ST/P002307/1,~ST/R002452/1,~ST/I006285/1 and ST/V005618/1,
STFC operations Grant No.~ST/R00689X/1. Computations were done on
the CSD3 and Fawcett (Cambridge), Cosma (Durham), Niagara (Toronto), Narval (Montreal),
Stampede2 (TACC) and Expanse (SDSC) clusters.

\bibliographystyle{iopart-num}
\bibliography{ref}

\end{document}